\documentclass[%
 reprint,
 amsmath,amssymb,
 aps,
]{revtex4-1}

\usepackage{graphicx}
\usepackage{dcolumn}
\usepackage{bm}

\begin{document}

\title{Soliton Mobility in Disordered Lattice}

\author{Zhi-Yuan Sun}
\email{sunzhiyuan137@aliyun.com}
\author{Shmuel Fishman}
\affiliation{Department of Physics, Technion-Israel Institute of
Technology, Haifa 320000, Israel }
\author{Avy Soffer}
\affiliation{Department of Mathematics, Rutgers University, New
Jersey 08854, U.S.A}

\date{\today}
\begin{abstract}
We investigate soliton mobility in the disordered Ablowitz-Ladik
(AL) model and the standard nonlinear Schr\"{o}dinger (NLS) lattice
with the help of an effective potential generalizing the
Peierls-Nabarro potential. This potential results from deviation
from integrability, resulting of randomness for the AL model, and of
both randomness and lattice discreteness for the NLS lattice.
Statistical properties of such a potential are analyzed, and it is
shown how the soliton mobility is affected by its size. The
usefulness of this effective potential in studying soliton dynamics
is demonstrated numerically. Further we propose two ways the soliton
transport in presence of disorder can be enhanced: one is to use
specific realizations of randomness, and the other one is to
consider a specific soliton pair.

\begin{description}
\item[PACS numbers]
05.45.Yv, 42.65.Tg
\end{description}
\end{abstract}

\maketitle

\textit{\textbf{Introduction}}---Discrete solitons (including
breathers---in our discussion referred as the localized modes of
soliton type but with temporarily oscillating profiles) for the
nonlinear Schr\"{o}dinger (NLS) type models on a lattice, have been
theoretically and experimentally studied in various areas of
physics, e.g., in nonlinear optics and solid-state physics
\cite{Lederer,Flach,Aubry,Kevrekidis}. These localized modes can
exist in the interplay of the nonlinearity and discreteness, and
their mobility is one of the most important aspects, for considering
the mass and energy transport in a lattice. One effective path to
physically understand mobility of the discrete soliton is using the
description in terms of the Peierls-Nabarro (PN) potentials
\cite{Kivshar,Johansson}. It could be understood as a type of
effective potential, resulting of deviation from integrability (we
will obtain similar deviation results of randomness for the
disordered AL model, and of both randomness and discreteness for the
disordered NLS lattice), with its amplitude being viewed as a
minimum barrier that the soliton should overcome to propagate in the
lattice. By virtue of such a concept, a series of works investigated
mobility of discrete solitons in various NLS lattices
\cite{Johansson,Hadzievski,Naether}.

One practical method to calculate the effective potential is
employing a collective coordinate approximation
\cite{Kivshar,Johansson}, with the assumption that the discrete
soliton propagates very slowly and adiabatically. In this potential
the soliton can be considered as an effective particle. Such
adiabatic approach provides a good approximate picture when the
radiation is negligible.

\textit{In the present paper}, we will study solitons for the AL
model and the standard NLS lattice in presence of disorder using
similar methods. Our investigations are relevant for experiments
using random media in optics \cite{Schwartz} and optical speckle
potentials in Bose-Einstein condensates \cite{Lye}. On the other
hand, nonlinearity and disorder are found to play important roles in
existence and stability of the localized excitation in a trimer
model \cite{Bai}. We will analyze statistical properties of the
generalized PN effective potential, and show how these affect the
soliton mobility. We will mainly focus on the relatively
short-time-scale soliton behavior, with the lattice soliton
initially injected with a small amount of kinetic energy (moving
slowly), and emitting negligible radiation. Under such situations,
certain basic features related to the PN effective potential are
believed to be revealed.

\begin{figure}[b]
\includegraphics[scale=0.21]{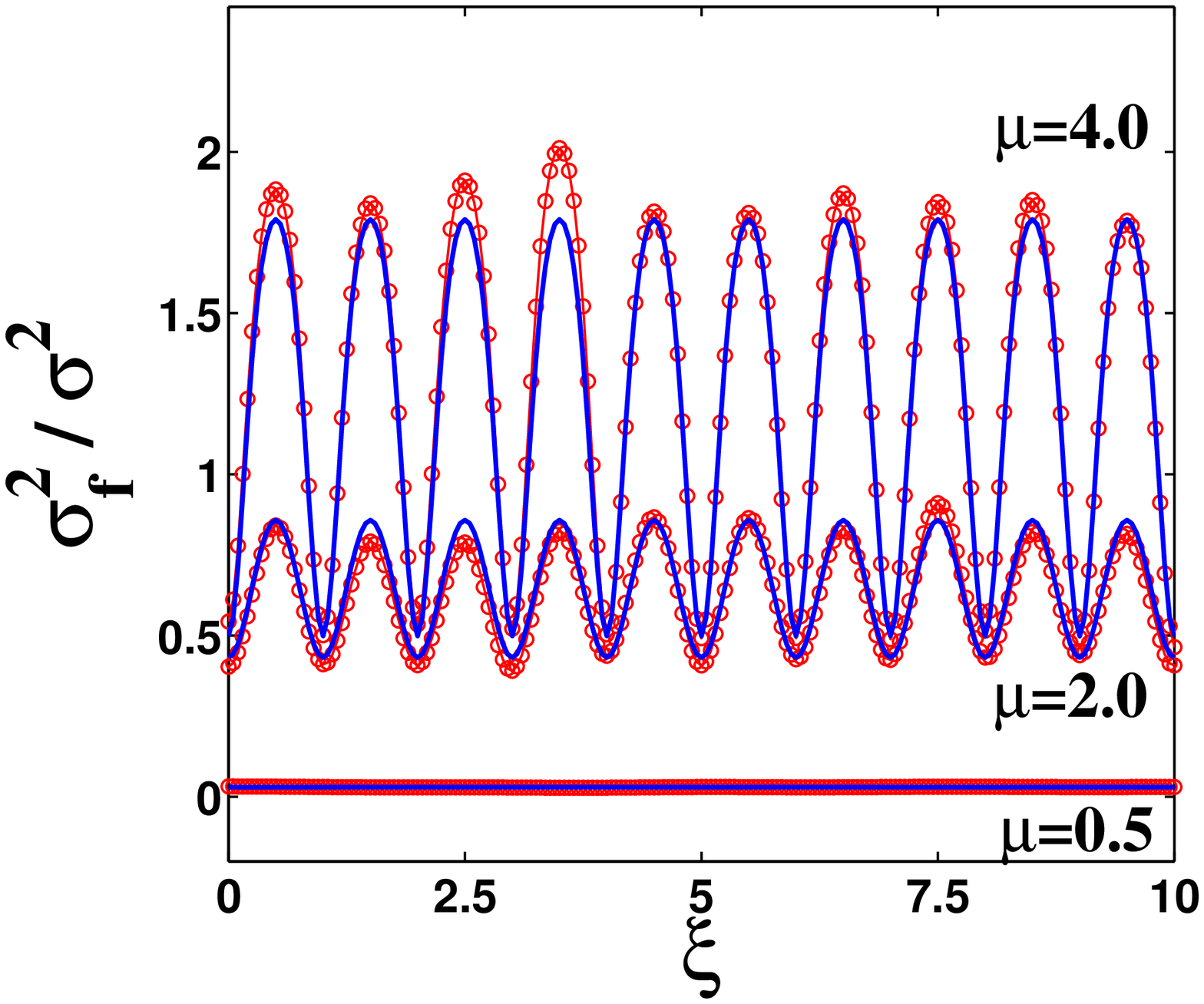}
\includegraphics[scale=0.21]{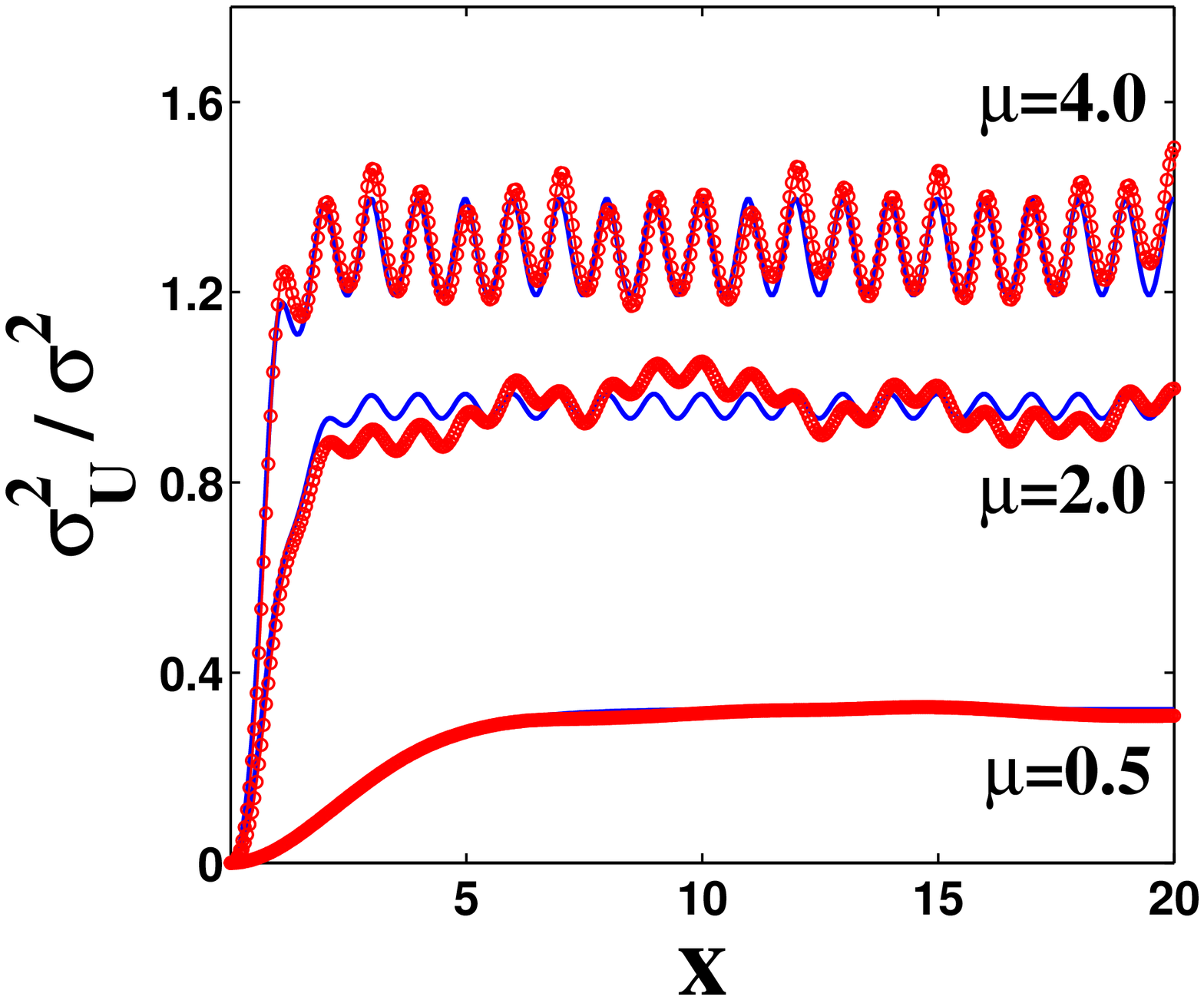}
\vspace{0cm} {\center\footnotesize\hspace{0cm}\textbf{(a)}
\hspace{3.5cm}\textbf{(b)}}\\
\includegraphics[scale=0.24]{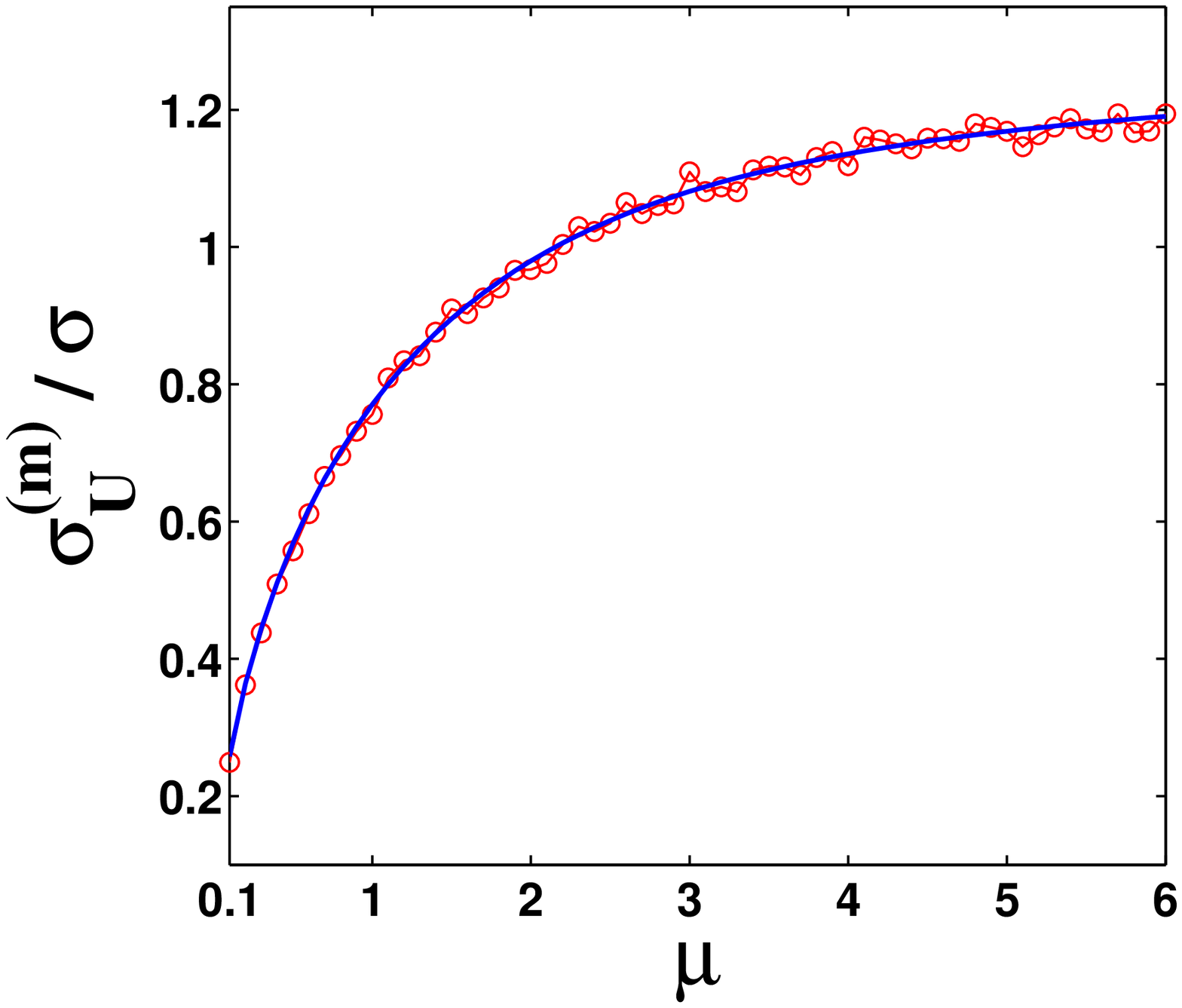}
\vspace{0cm} {\center\footnotesize\hspace{0cm}\textbf{(c)}}
\caption{\label{A} (Color online) Statistical properties of the
equivalent force $f(\xi)$ and effective potential $U(x)$. The blue
solid lines represent the results from (\ref{7}) or (\ref{8.1}), and
the red circles represent the statistically simulated results. In
simulation, we discretize the coordinate with $\Delta\xi=0.05$, and
the variance is computed for $10^3$ realizations of random
potentials. (a) Comparison between the statistically simulated
variance $\sigma_f^2/\sigma^2$ and (\ref{7}) for $\mu=0.5$, $2.0$,
and $4.0$ ($\sigma=0.02$). (b) Comparison between the statistically
simulated variance $\sigma_U^2/\sigma^2$ and (\ref{8.1}). (c)
$\sigma_U^{(m)}/\sigma$ of (\ref{8.2}) as a function of $\mu$.}
\end{figure}

\textit{\textbf{Soliton mobility in Ablowitz-Ladik lattice with
disorder}}---We start from the integrable Ablowitz-Ladik (AL) model
\cite{Ablowitz}, with a random potential term, defined by
\begin{equation}
i \dot{\psi_n} = -(\psi_{n-1} + \psi_{n+1})(1 + |\psi_n|^2) +
\varepsilon_n \psi_n~,\label{1}
\end{equation}
where $\psi_n$ is the wavefunction at site $n$ and time $t$, while
$\varepsilon_n$ is the normally distributed random potential
(uncorrelated) with zero mean value and standard variance $\sigma$,
that is, $\langle\varepsilon_n\rangle=0$ and $\langle\varepsilon_n
\varepsilon_{n'}\rangle=\sigma^2\delta(n-n')$. Integrability of the
AL model in absence of the potential leads its PN potential barrier
to vanish \cite{Kivshar} (as well as vanishing of the resonances
that generate radiation \cite{Flach}), and admits an exact mobile
soliton solution at arbitrary velocity, written as
\begin{equation}
\psi_n(t) = \frac{\sinh(\mu)}{\cosh[\mu(n-x)]} \exp[i k (n-x) + i
\alpha]~, \label{2}
\end{equation}
where (without disorder) the time-dependent parameters $x$ and
$\alpha$ can be expressed as $\dot{x} = 2 \frac{\sinh(\mu)}{\mu}
\sin(k)$ and $\dot{\alpha} = 2 [\cosh(\mu)\cos(k) + \frac{k}{\mu}
\sinh(\mu)\sin(k)]$. We now assume the random potential is weak and
the soliton velocity is slow, then apply the adiabatic approximation
\cite{Vakhnenko,Cai} to derive the evolution equations for the
soliton parameters $x$ and $k$ with disorder (assuming
$\dot{\mu}=0$)
\begin{subequations}\label{3}
\begin{align}
&\hspace{0mm} \dot{x} = \frac{2\sinh(\mu)}{\mu} \sin(k)~,\label{3a}\\
&\hspace{0mm} \dot{k} = -\sum^{+\infty}_{n=-\infty}
\frac{\varepsilon_n \sinh^2(\mu) \tanh[\mu (n-x)]}{\cosh[\mu
(n+1-x)] \cosh[\mu (n-1-x)]}~. \label{3b}
\end{align}
\end{subequations}
For the system (\ref{3}), the effective potential, regarded as the
PN potential, can be written as
\begin{equation}
U(x)=-\int_0^x f(\xi)d\xi~,\label{3.5}
\end{equation}
with the equivalent \textit{force} $f(\xi) = \sum \varepsilon_n
\phi(n-\xi)$, where $\phi(z)= -\frac{\sinh^2(\mu) \tanh(\mu
z)}{\cosh[\mu (z+1)] \cosh[\mu(z-1)]}$. The force $f(\xi)$ also has
zero mean value, and its correlation function $r_f$ can be derived
due to its linear property,
\begin{equation}
r_f(\xi,\xi') = \langle f(\xi)f(\xi') \rangle  = \sigma^2
\sum^{+\infty}_{n=-\infty} \phi(n-\xi)\phi(n-\xi')~.\label{4}
\end{equation}
Using the Poisson summation formula and residue theorem, we can
evaluate the sums in (\ref{4}), and obtain
\begin{equation}
r_f(\xi,\xi') = \sigma^2 \left[ \mathfrak{R}_0(\mu,\Delta
\xi)+2\sum_{s=1}^{+\infty} \mathfrak{R}_s(\mu,\xi,\Delta \xi)
\right]~,\label{5}
\end{equation}
where ($\Delta\xi=\xi-\xi'$),
\begin{widetext}
\begin{subequations}\label{6}
\begin{eqnarray}
\mathfrak{R}_0 =&& \sinh^2(\mu)\{ 2\Delta\xi \coth(\mu\Delta\xi)
\textrm{csch}[\mu(\Delta\xi-1)] \textrm{csch}[\mu(\Delta\xi+1)] -
(\Delta \xi-2) \coth[\mu(\Delta\xi-1)]
\textrm{csch}[\mu(\Delta\xi-2)] \textrm{csch}(\mu\Delta\xi) \nonumber\\
&& - (\Delta \xi+2) \coth[\mu(\Delta\xi+1)]
\textrm{csch}[\mu(\Delta\xi+2)]
\textrm{csch}(\mu\Delta\xi)\}~,\label{6a}
\end{eqnarray}
\begin{eqnarray}
\mathfrak{R}_s = &&\frac{\pi\sinh^2(\mu)}{\mu} \textrm{csch}
\left(\frac{\pi^2 s}{\mu}\right) \cos[\pi s (2\xi-\Delta\xi)] \{2
\coth(\mu\Delta\xi) \textrm{csch}[\mu(\Delta\xi+1)]
\textrm{csch}[\mu(\Delta\xi-1)]
\sin(\pi s \Delta\xi) \nonumber \\
&&-\coth[\mu(\Delta\xi-1)] \textrm{csch}(\mu\Delta\xi)
\textrm{csch}[\mu(\Delta\xi-2)] \sin[\pi s (\Delta\xi-2)] \nonumber\\
&&-\coth[\mu(\Delta\xi+1)] \textrm{csch}(\mu\Delta\xi)
\textrm{csch}[\mu(\Delta\xi+2)] \sin[\pi s
(\Delta\xi+2)]\}~.\label{6b}
\end{eqnarray}
\end{subequations}
\end{widetext}
Eqs.~(\ref{5}) and (\ref{6}) apparently show that, due to
discreteness, $f(\xi)$ is a nonstationary random process (depending
not only on $\Delta\xi$) \cite{Papoulis}. With (\ref{5}), we derive
the variance of $f(\xi)$ as
\begin{widetext}
\begin{equation}
\sigma_f^2(\xi)/\sigma^2 = r_f(\xi'\rightarrow\xi)/\sigma^2 =
3\coth(\mu)-\tanh(\mu)-\frac{3}{\mu} - \frac{6\pi^2}{\mu^2}
\sum_{s=1}^{+\infty} s~ \textrm{csch}
\left(\frac{\pi^2s}{\mu}\right) \cos(2\pi s \xi)~.\label{7}
\end{equation}
\end{widetext}
Generally speaking, when $\mu\ll1$, $f(\xi)$ can be approximately
seen as a stationary random process, with
$\sigma_f^2/\sigma^2\approx3\coth(\mu)-\tanh(\mu)-\frac{3}{\mu}$
(when $\mu\lesssim0.3$, it accords well with the continuous limit
result $\sigma_f^2/\sigma^2 =
\frac{4\sinh^4(\mu)}{15\mu}\approx\frac{4}{15}\mu^3$); when
$1\lesssim\mu\lesssim3$, $f(\xi)$ is a nonstationary random process
with a periodic variance (the terms for $s\geqslant2$ are neglected
as small terms)
$\sigma_f^2(\xi)/\sigma^2\approx3\coth(\mu)-\tanh(\mu)-\frac{3}{\mu}-
\frac{6\pi^2}{\mu^2} \textrm{csch} \left(\frac{\pi^2}{\mu}\right)
\cos(2\pi \xi)$; when $\mu\gtrsim3$, we may have to consider overlap
of the terms with $s\geqslant2$. We numerically calculate the force
$f(\xi)$ for a large number of realizations of the random potential
to derive its variance, and make a comparison with (\ref{7}). Good
agreement can be seen in Fig.~\ref{A}(a), for two different regimes.

With the statistical property of the force $f(\xi)$, we next
consider the effective potential $U(x)$. Apparently, it is a
nonstationary random process with zero mean value, and its
correlation function can be derived as
\begin{equation}
r_U(x,x') = \langle U(x)U(x') \rangle = \int_0^x \int_0^{x'}
r_f(\xi,\xi') d\xi d\xi'~.\label{8}
\end{equation}
We can numerically integrate (\ref{8}) with appropriate truncation
of (\ref{5}) according to the value of $\mu$, and further obtain the
variance of $U(x)$ as
\begin{equation}
\sigma_U^2(x)/\sigma^2 = r_U(x'\rightarrow x)/\sigma^2~.\label{8.1}
\end{equation}
Three typical examples of (\ref{8.1}), compared to numerical results
found sampling various realizations using (\ref{3.5}), are shown in
Fig.~\ref{A}(b). We can see that, after an increase in short
distance, $\sigma_U/\sigma$ approaches a periodic type function (for
small enough $\mu$, the periodicity can be neglected, e.g.,
$\mu=0.5$) with a nonzero mean value (averaged in finite periods
along $x$). We denote such mean value as $\sigma_U^{(m)}/\sigma$,
where $\sigma_U^{(m)}$ is theoretically defined as
\begin{equation}
\sigma_U^{(m)} = \lim_{L\rightarrow\infty} \frac{1}{L} \int_0^L
\sigma_U(x)dx~,\label{8.2}
\end{equation}
and further compute that the amplitude of the periodic function is
not more than $6\%$ of $\sigma_U^{(m)}/\sigma$, even up to $\mu=6$.
Thus, $\sigma_U^{(m)}$ can be employed, in the sense of statistics,
as the PN-type potential barrier that relates the soliton mobility
in presence of disorder. $\sigma_U^{(m)}/\sigma$ as a function of
$\mu$, is presented in Fig.~\ref{A}(c), as well as the comparison
with numerical results for multiple realizations of the random
potentials. Generally speaking, the smaller soliton has a larger
mobility due to its smaller effective potential barrier, however,
there exist approximately two regimes: one is the \textit{sharp
slope} regime ($\mu\lesssim1$) where $\sigma_U^{(m)}/\sigma$
decreases quickly with $\mu$ decreasing; the other one is the
\textit{flat slope} regime ($\mu\gtrsim3$) where
$\sigma_U^{(m)}/\sigma$ decreases considerably slowly with $\mu$
decreasing.

\begin{figure}[b]
\includegraphics[scale=0.22]{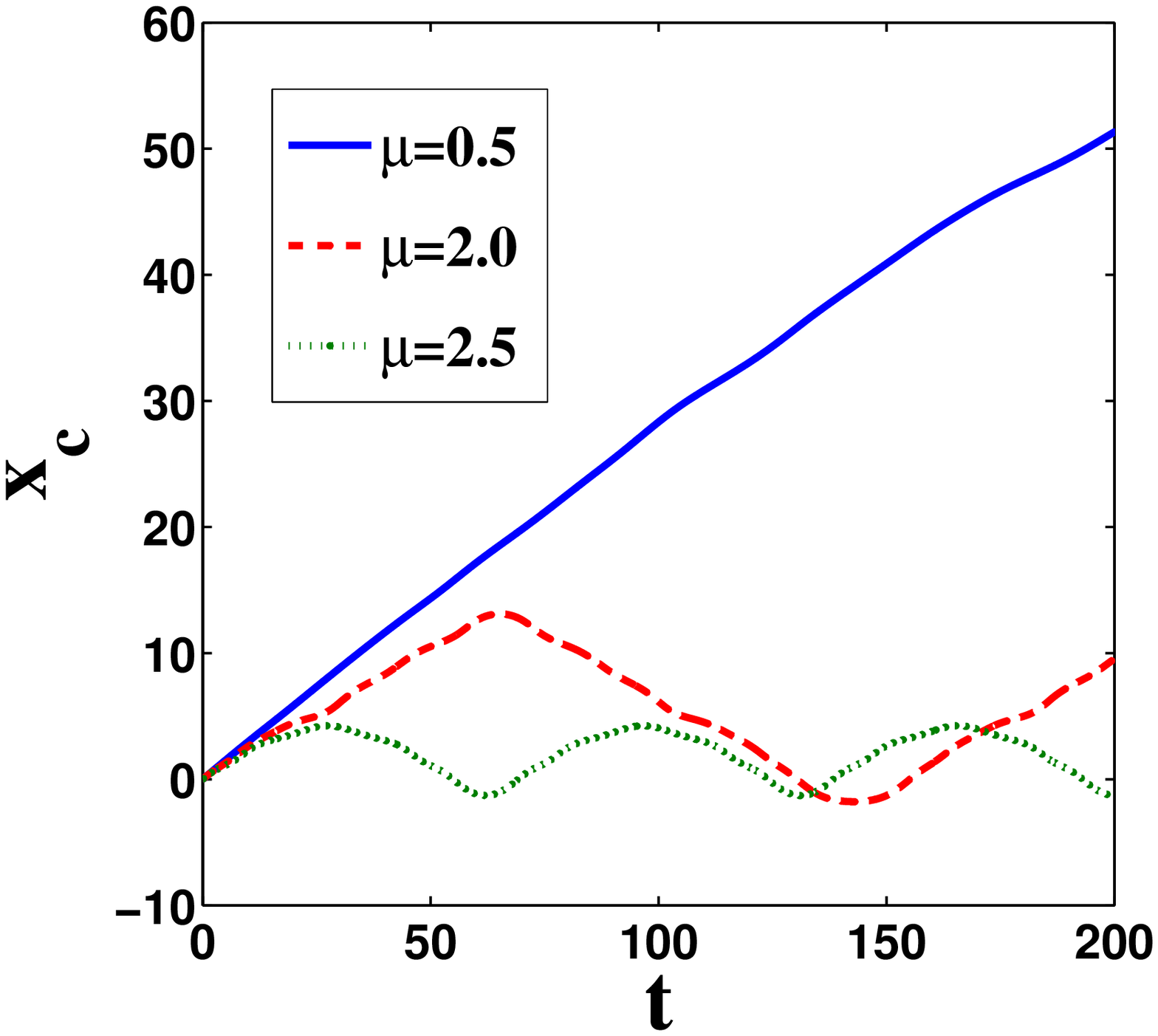}
\includegraphics[scale=0.22]{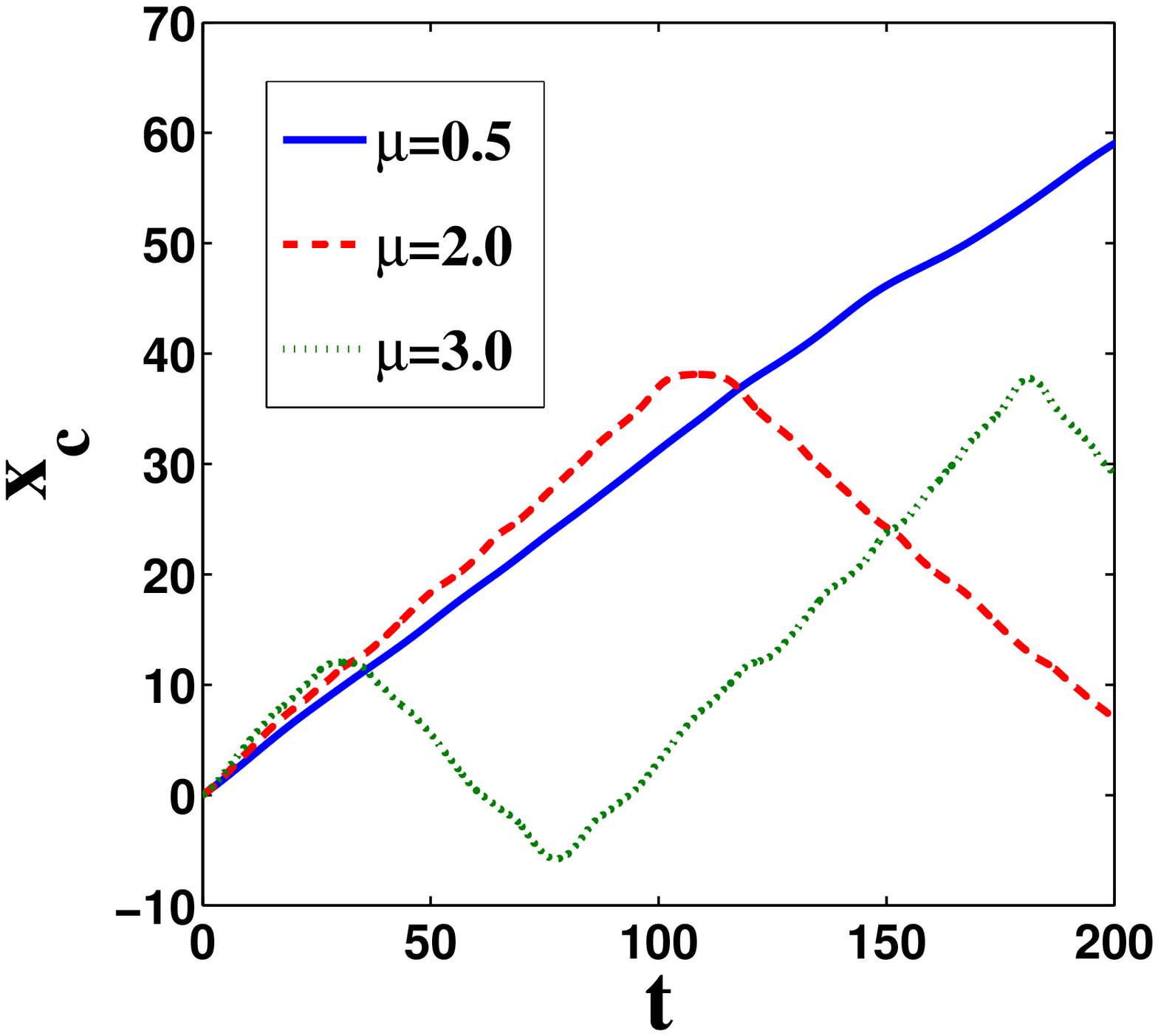}
\vspace{0cm} {\center\footnotesize\hspace{0cm}\textbf{(a)}
\hspace{3.5cm}\textbf{(b)}} \caption{\label{B} (Color online)
Soliton center of mass $x_c$ as a function of time $t$ for two
typical realizations of random potentials ($\sigma=0.01$ and
$v_0=0.3$), found from (\ref{1}). }
\end{figure}

To check such mobility, we directly proposed numerical simulation of
Eq.~(\ref{1}), with the soliton solution (\ref{2}) as the initial
condition. The soliton behavior was observed for more than $100$
realizations of random potentials with $\sigma=0.01$, and the
soliton was given a small initial velocity $v_0=0.3$ for each
realization. Two typical results, with the soliton center of mass
$x_c$ as function of $t$, are presented in Figs.~\ref{B}(a) and (b).
We can see that, no matter whether the soliton decelerates or
accelerates (on average) for some time, the larger soliton is
reflected or trapped by the randomness stronger than the small
soliton. In principle, the soliton reflects near some large
effective potential barrier, or gets trapped between two large
barriers. Before the first reflection, whether deceleration or
acceleration occurs, is depending on the details of the random
potential. Generally, we can approximately use a time difference to
identify such deceleration (acceleration) stage, defined as
\begin{equation}
\tau(x)=\int_0^x dx \left[\frac{\mu}{2\sinh(\mu)\sin(k)} -
\frac{1}{v_0}\right]~,\label{9}
\end{equation}
where $\cos(k) = \frac{\mu}{2\sinh(\mu)}U(x) + \cos(k_0)$, as
derived from the conservation of the effective total energy
$H=-\frac{2\sinh(\mu)}{\mu}\cos(k)+U(x)$. Thus, if $\tau<0$, there
exists an averaged acceleration process within $[0,x]$ (before first
reflection); while $\tau>0$, it is an averaged deceleration process.
One interesting idea is to use the region where randomness
accelerates the soliton in finite time. For instance, as seen in
Fig.~\ref{B}(b), for relatively large soliton, we may choose the
random potential sections before the first reflection point, and
arrange them periodically, to realize the mass transport in one
direction for some time.

\begin{figure}[b]
\includegraphics[scale=0.225]{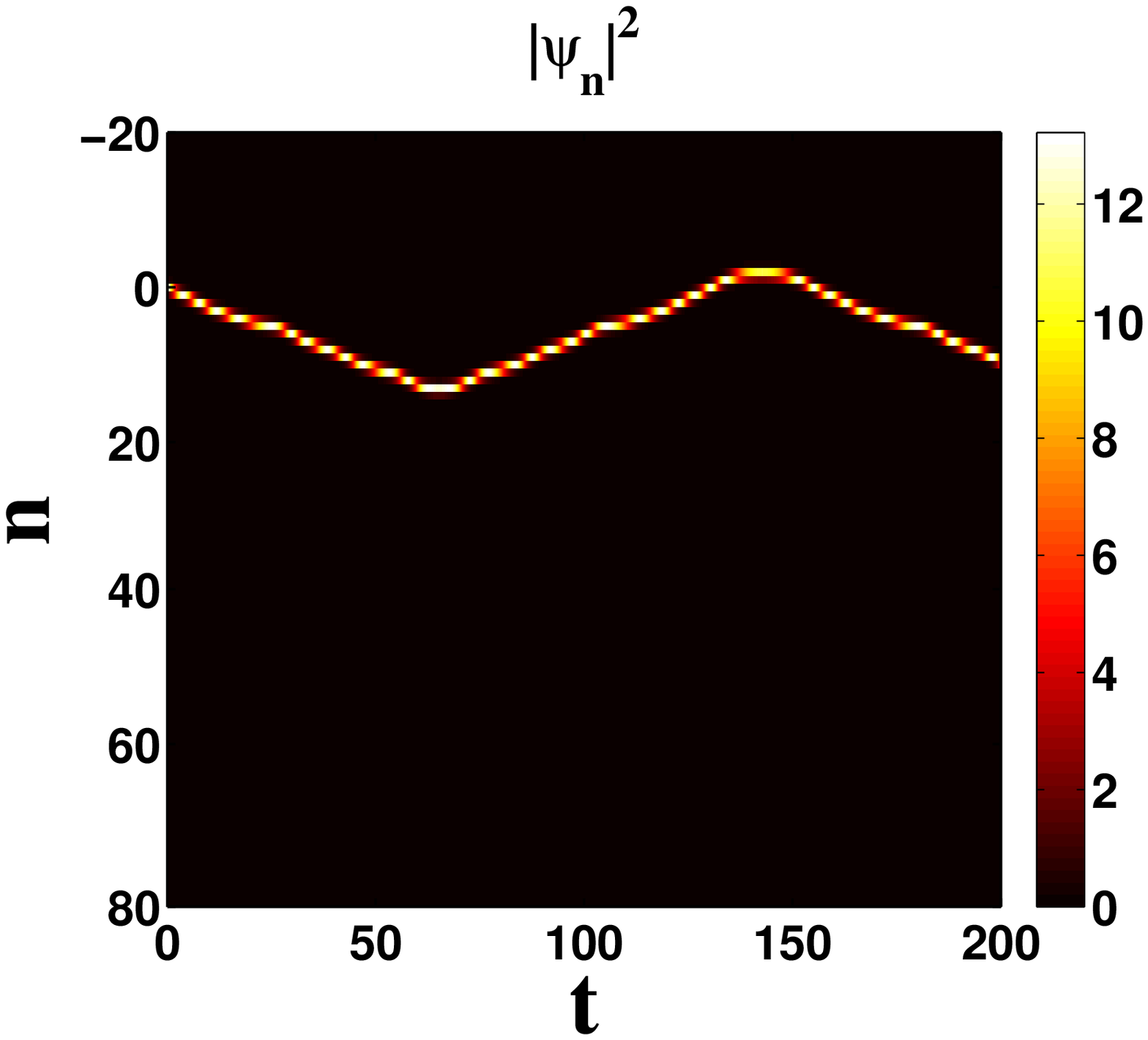}
\includegraphics[scale=0.225]{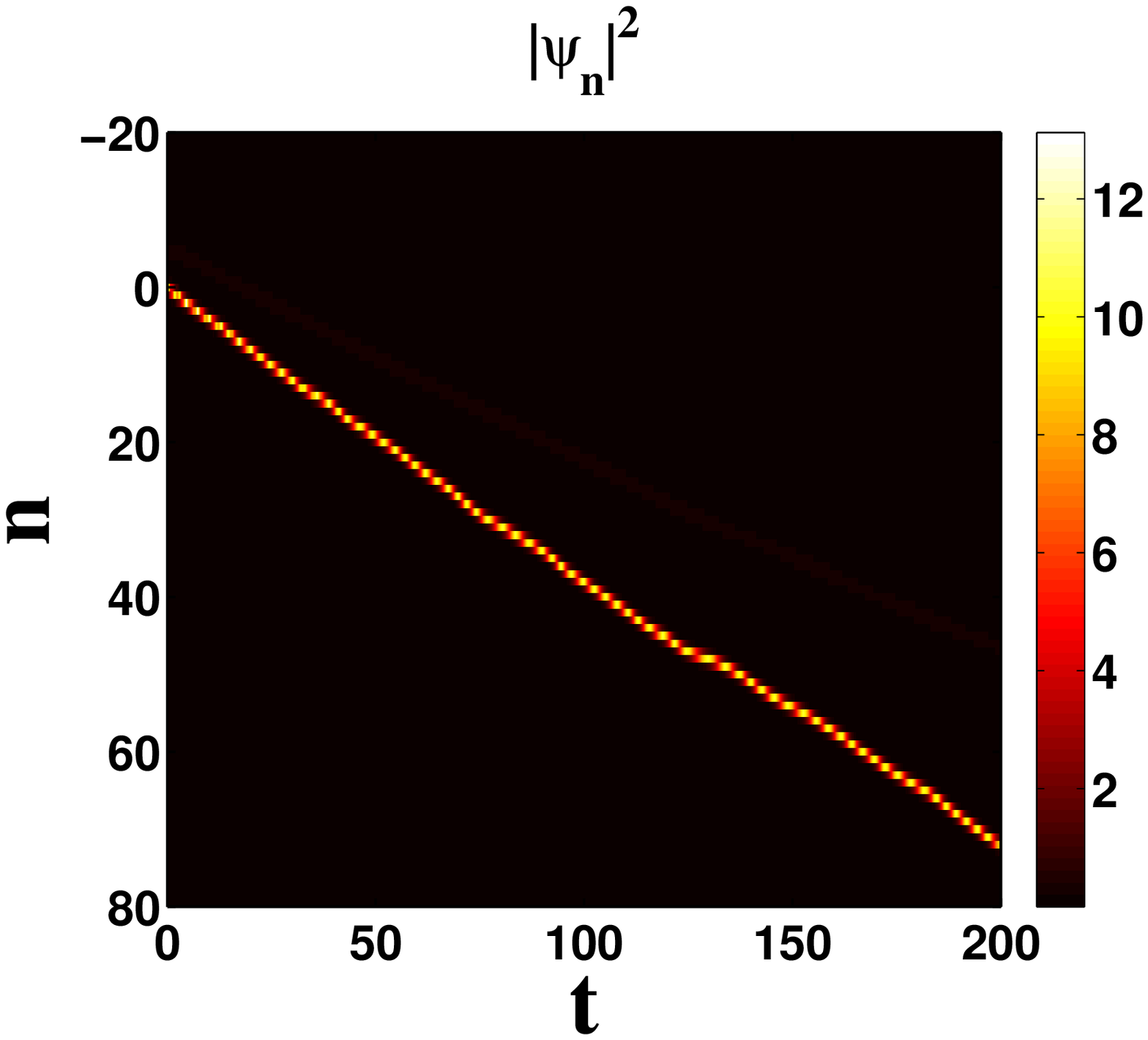}
\vspace{0cm} {\center\footnotesize\hspace{0cm}\textbf{(a)}
\hspace{3.5cm}\textbf{(b)}} \caption{\label{C} (Color online)
Intensity map of the soliton propagation in a realization of random
potential with $\sigma=0.01$. (a) A single soliton with $\mu=2$ and
$v_0=0.3$. (b) A pair of solitons with $\mu_1=2$, $\mu_2=0.5$, and
$v_0=0.3$. The initial separation distance $\Delta x$ and phase
difference $\Delta \alpha$ are $5$ and $\pi$, respectively. }
\end{figure}

Another aspect is to consider a pair of solitons, where one is large
and the other one is small, with the same initial velocity and a
short separation distance between them (may be partially
overlapped). In order to enhance transportation of the large
soliton, we arrange the small one to follow the large one, and give
them an initial phase difference $\pi$ that provides a repulsive
interaction between both solitons. When the large soliton encounters
a large effective potential barrier that would obstruct its
mobility, while the small one might be still well propagating due to
its larger mobility, and is expected to \textit{push} the large one
to \emph{help} it overcome the barrier. A similar mechanism to
trigger a migration by a low-amplitude solitary wave that collides
with an excitation has been described for a pure NLS lattice
\cite{Rumpf}. As an example, such idea is realized and illustrated
in Fig.~\ref{C} for a specific realization of randomness.
Fig.~\ref{C}(a) presents a large single soliton ($\mu=2$) trapped by
the random potential, while in Fig.~\ref{C}(b), we add the auxiliary
small soliton ($\mu=0.5$) with the separation distance of $5$. It
shows that the large soliton can now propagate without reflection or
trapping for some time. The interaction of the small soliton
($\mu=0.5$), with the large one results in the motion of the large
soliton in the same direction without reflection. Similar results
were found also for other realizations of the disorder.

\begin{figure}[b]
\includegraphics[scale=0.255]{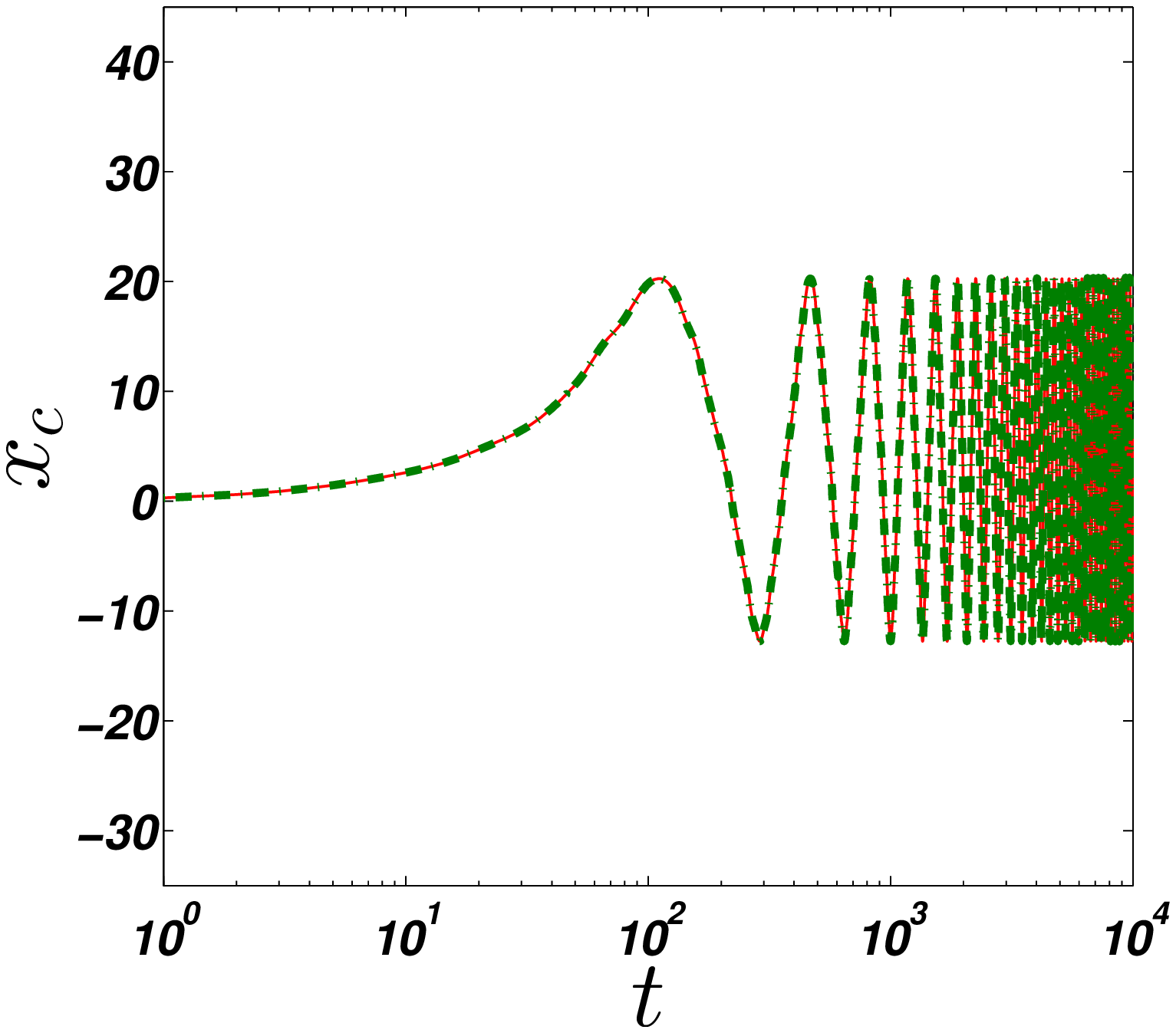}
\includegraphics[scale=0.255]{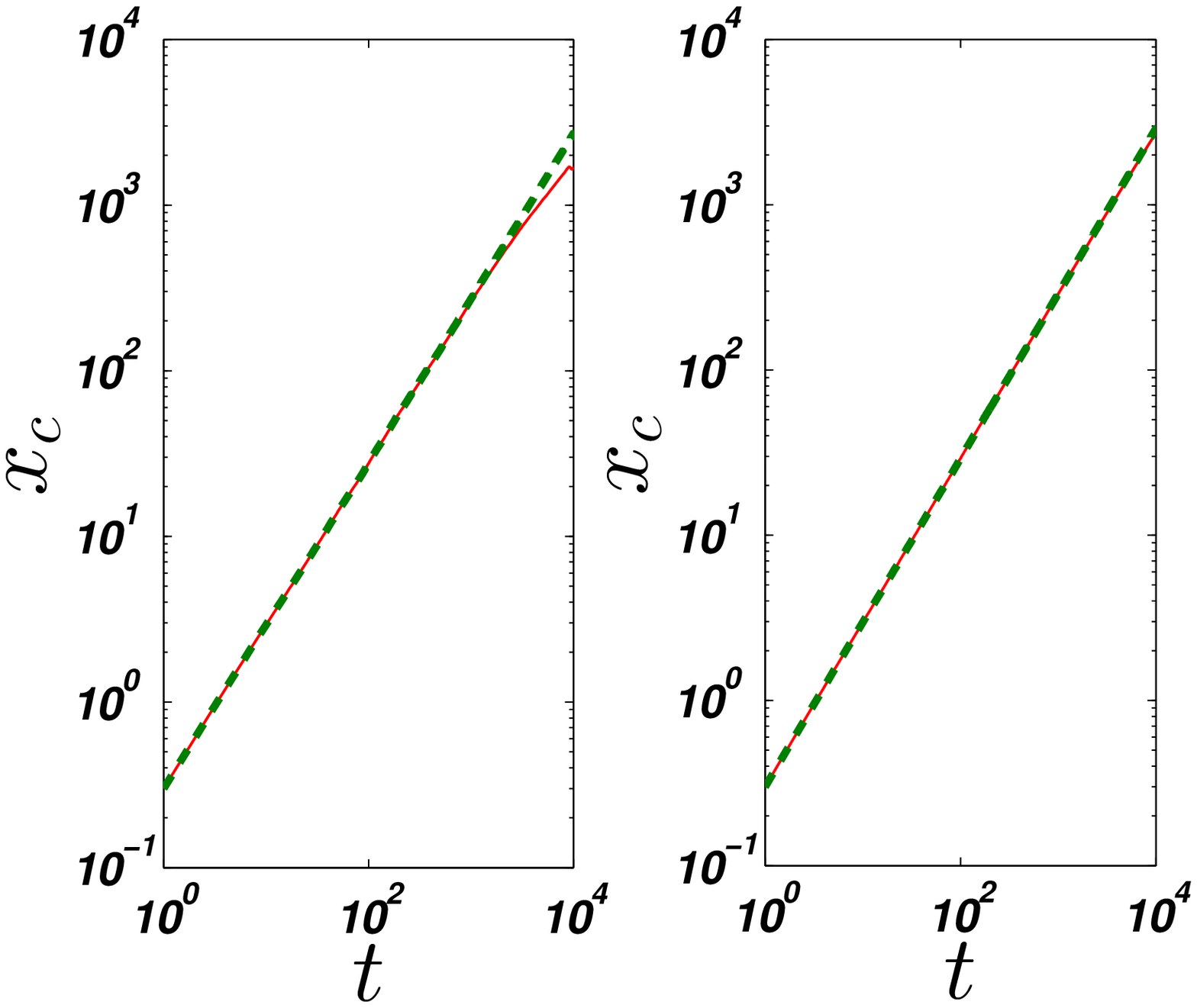}
\vspace{0cm} {\center\footnotesize\hspace{0cm}\textbf{(a)}
\hspace{3.5cm}\textbf{(b)}} \caption{\label{X} (Color online)
Comparison of the soliton trajectory $x_c$ obtained by numerically
integrating Eq.~(\ref{1}) (red solid line) and effective potential
approach using $U(x)$ of (\ref{3.5}) (green dashed line). (a) The
random potential $\varepsilon$ used is of $\sigma=0.01$, and the
soliton parameters are $\mu=1.0$ and $v_0=0.3$. (b) Left Panel: the
random potential is $\varepsilon{'}=0.5\varepsilon$; Right Panel:
$\varepsilon{'}=0.2\varepsilon$. The soliton parameters are
$\mu=0.5$ and $v_0=0.3$.  }
\end{figure}

For the large soliton, the randomness-induced radiation is still
very small for considerably long time, and the soliton behavior can
be well described by the effective potential [ a typical comparison
is shown in Fig.~\ref{X}(a) up to $t=10^4$ ]. Generally speaking,
the large soliton is apt to be trapped by the randomness after some
sequences of deceleration and acceleration periods (the soliton
behavior was observed in the long-time simulation for more than 10
realizations of the random potential). For the small soliton, if the
randomness is weak enough, the effective potential also gives a good
approximation for long time. In Fig.~\ref{X}(b), we decrease the
strength of the random potential from the left to the right panels,
and show that the effective potential approach agrees better with
the numerical solutions of Eq.~(\ref{1}) for weaker randomness. On
the other hand, if the randomness is relatively strong, the small
soliton was observed to continue radiating its mass and kinetic
energy during the long-time propagation, and even exhibit visible
deformation on its profile. Such a case may be analyzed using the
method of modulation equation \cite{Soffer}, which is not within the
scope of this paper.

\textit{\textbf{Behavior for the standard NLS lattice with
disorder}}---We study the standard NLS lattice
\begin{equation}
i \dot{\psi_n} = -(\psi_{n-1} + \psi_{n+1}) - \nu |\psi_n|^2 \psi_n
+ \varepsilon_n \psi_n~.\label{10}
\end{equation}
We consider this equation as the AL model with a perturbation term
on the RHS, $R_n = |\psi_n|^2(\psi_{n+1} + \psi_{n-1} -\nu\psi_n) +
\varepsilon_n \psi_n$, and use the adiabatic approximation to derive
the similar generalized PN effective potential when $\mu$ is not too
large ($\mu\ll3$),
\begin{equation}
\mathcal{U}(x) =
-\frac{2\pi^2\nu\sinh^2(\mu)}{\mu^3\sinh\left(\frac{\pi^2}{\mu}\right)}
\cos(2\pi x) + U(x)~,\label{11}
\end{equation}
where $\dot{\mu}=0$ and
$\dot{x}\approx2\sinh(\mu)\tanh(\mu)\sin(k)/\mu^2$, and $U(x)$ is
the same randomness generated effective potential (\ref{3.5}). In
the limit $\mu\rightarrow0$, the soliton is wide, therefore the
discreteness of the lattice is not important, and the result of the
continuous integrable model is approached. Naturally, we may use the
following parameter to approximately decide which factor dominates,
\begin{equation}
\kappa = \sigma_U^{(m)} /
\frac{2\pi^2\nu\sinh^2(\mu)}{\mu^3\sinh\left(\frac{\pi^2}{\mu}\right)}
= \frac{\sigma}{\nu} \left[\frac{\mu^3
\sinh\left(\frac{\pi^2}{\mu}\right)
\lambda(\mu)}{2\pi^2\sinh^2(\mu)} \right]~,\label{12}
\end{equation}
where $\lambda(\mu)$ denotes the curve presented in Fig.~\ref{A}(c).
For the typical parameters $\sigma=0.01$ and $\nu=1$, when
$\mu\ll1$, $\kappa\gg1$, the randomness dominates the soliton
behavior; when $\mu\approx1$, $\kappa\approx2.6$, these two
potentials are of the same magnitude; when $\mu\gg1$, $\kappa\ll1$,
the influence of randomness can be ignored. And, if the soliton is
not too large, influence of the randomness on the soliton mobility,
with regard to its size, is similar to that found for the model
(\ref{1}).

\begin{figure}[b]
\includegraphics[scale=0.25]{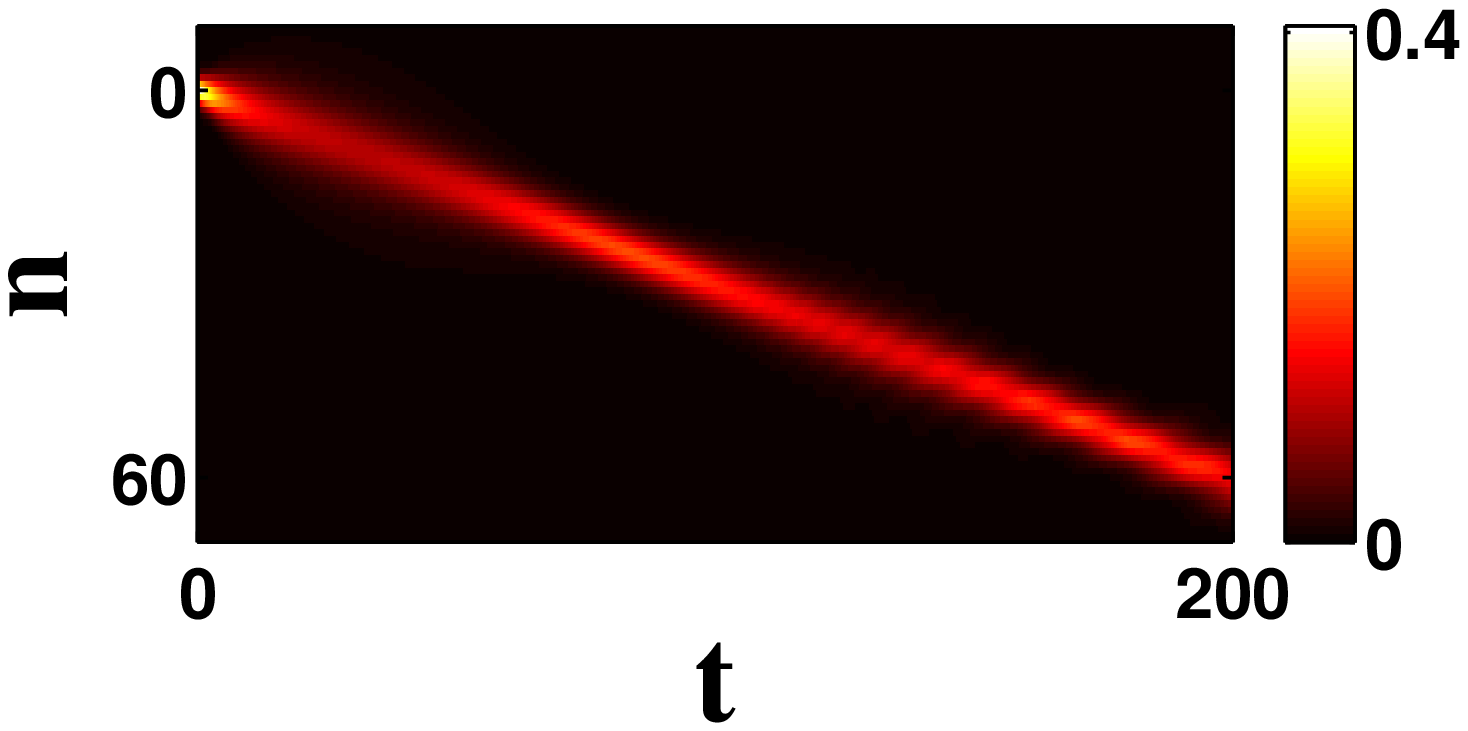}
\includegraphics[scale=0.25]{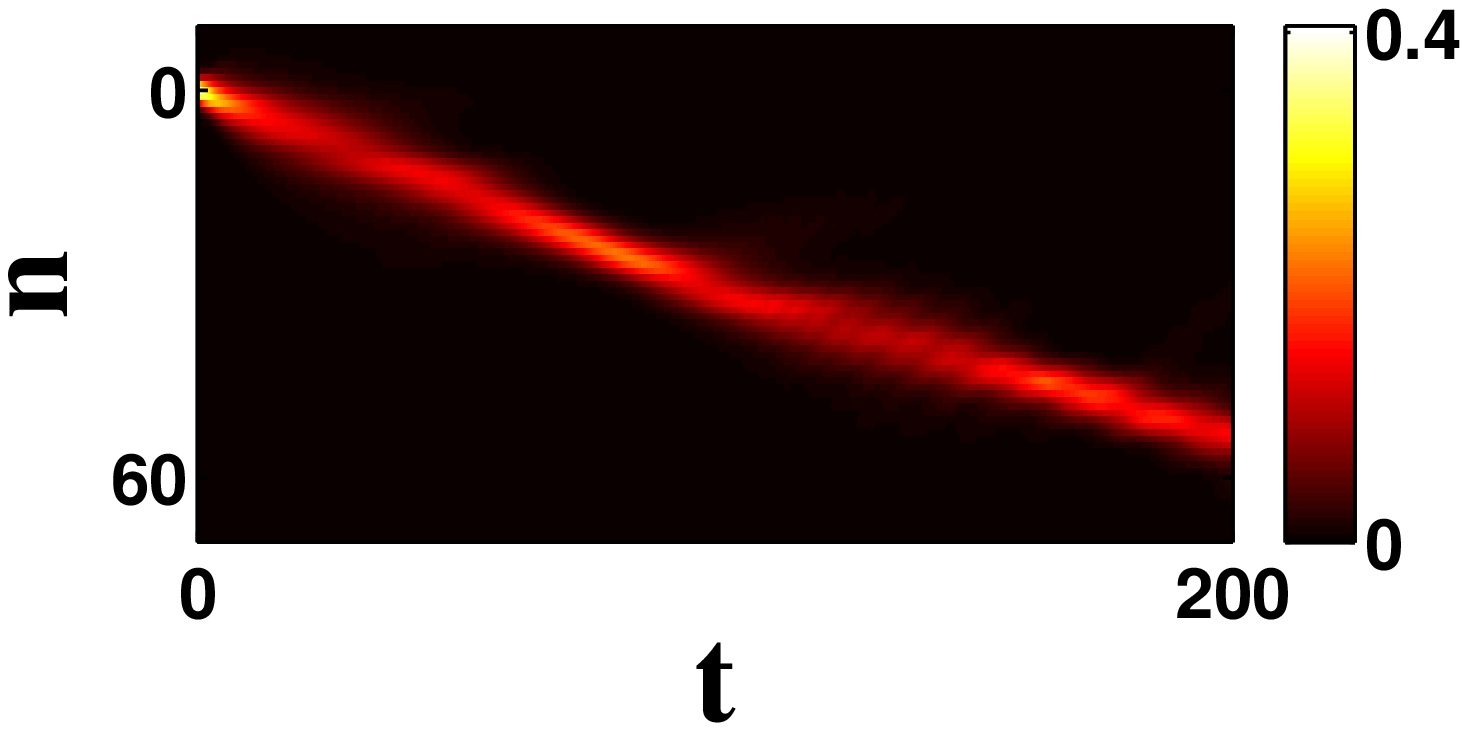}
\\ \hspace{0.1mm}
\includegraphics[scale=0.25]{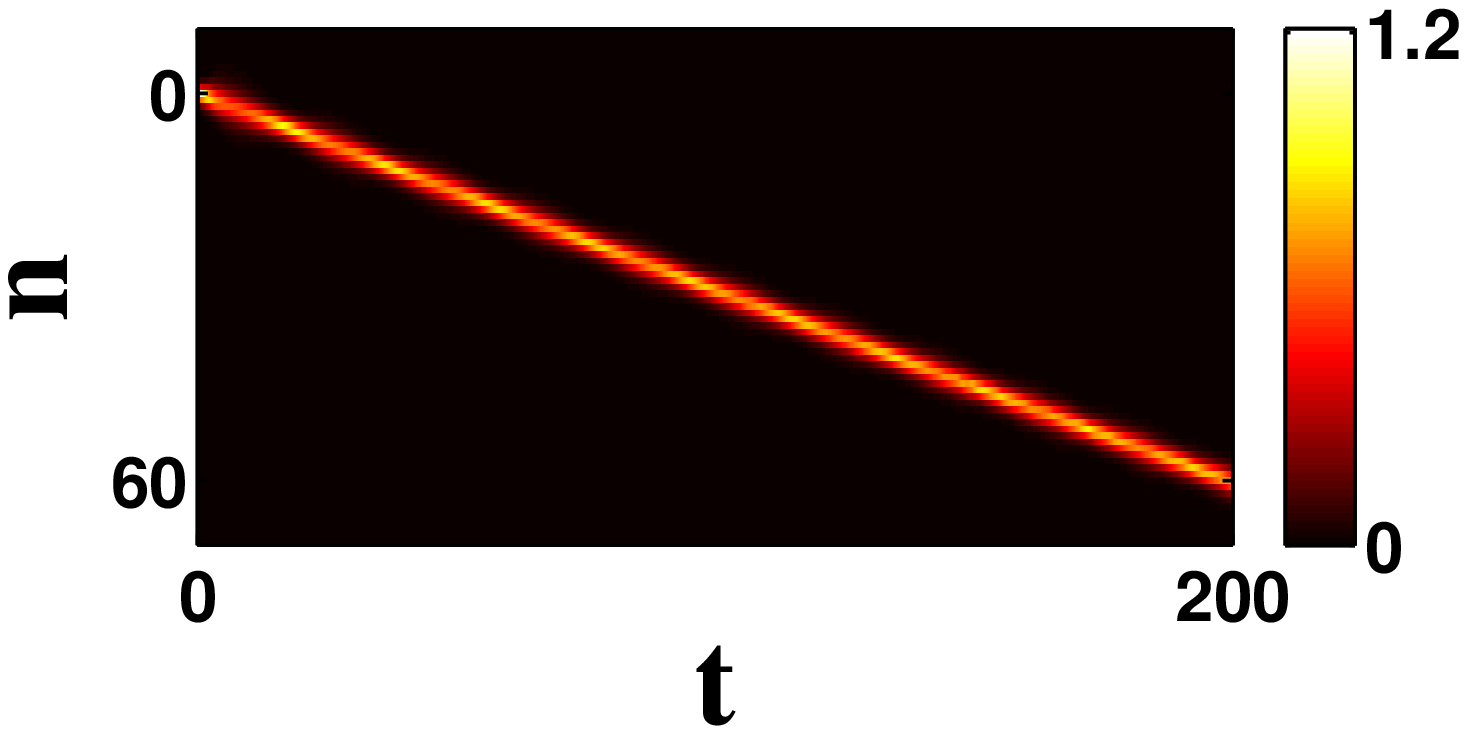}
\includegraphics[scale=0.25]{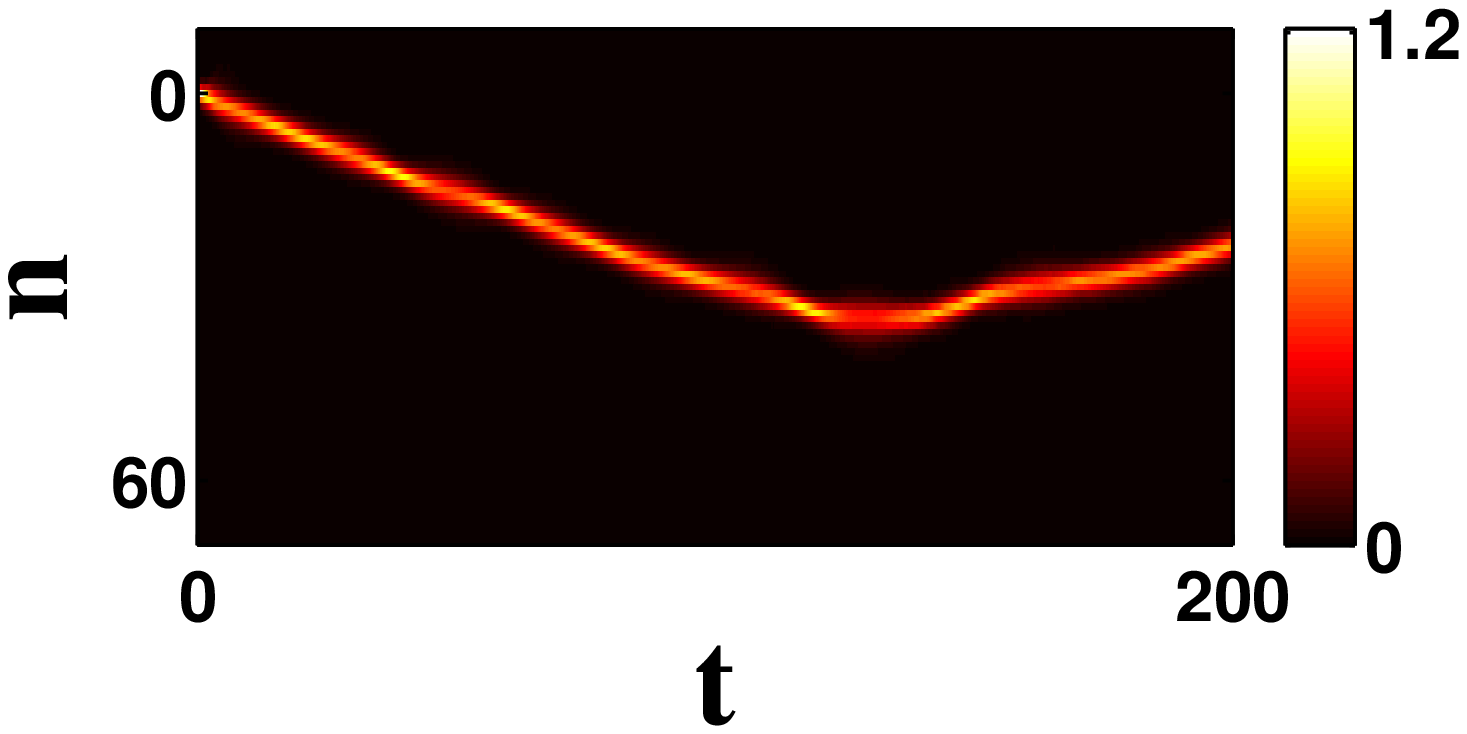}
\vspace{0cm}
{\center\footnotesize\hspace{0cm}\textbf{(a)}\hspace{3.6cm}\textbf{(b)}}
\caption{\label{D} (Color online) (a) Intensity $|\psi_n|^2$ for two
generated breathers of the NLS lattice (\ref{10})
($\varepsilon_n=0$) with almost the same velocity $v\approx0.30$.
The initial conditions are chosen as (\ref{2}) with $\mu=0.60$ and
$k=0.150$ for the upper panel, while $\mu=0.95$ and $k=0.178$ for
the lower panel ($\nu=0.9$). (b) The two breathers are propagating
in presence of the random potential with $\sigma=0.018$. }
\end{figure}

Here we make some comments: in fact, if the soliton is very small,
it is easy to emit relatively strong radiation induced by the
randomness, even to be greatly destroyed on its profile. Such a
condition may be remarkably out of the adiabatic approximation. If
the soliton is very large, its mobility becomes much smaller, due to
the large potential barrier of discreteness. On the other hand,
strictly speaking, the \textit{soliton} discussed here should be
replaced by the \textit{breather} ($\dot{\mu}\neq0$), however, for
some parameters, our study may generally and effectively provides a
physical description of the lattice soliton mobility with disorder.
As an example, we give a set of simulation results in Fig.~\ref{D}.
We know that, Eq.~(\ref{10}) with $\varepsilon_n=0$ has no exact
mobile soliton (breather) solutions, however, we could start from a
sech type soliton, after initially emitting a small part of
radiation, to numerically generate an approximate breather solution
that can propagate for considerably long time \cite{Franzosi}. In
Fig.~\ref{D}(a), we generate such two breathers, one is small (the
upper panel), the other one is large (the lower panel), with almost
the same small velocity. Then, we add a realization of the random
potential, and show the simulation result in Fig.~\ref{D}(b).
Apparently, we can see that, the mobility of the large breather is
obstructed by the randomness, while the small one is nearly not
affected, which is the similar feature as shown in Fig.~\ref{B}.

\textit{\textbf{Conclusions}}---With help of the generalized PN
effective potential, we investigated mobility of lattice solitons in
presence of disorder. We analyzed methods to enhance the mobility of
solitons of the AL model (\ref{1}) in a weak random potential. In
some situations (weak randomness and short time), these can be
considered as particles moving in the effective potential $U(x)$ of
(\ref{3.5}). We find two ways to enhance mobility: (a) introducing a
random potential that is engineered so that intervals of the random
potential where acceleration takes place appear one after the other;
(b) a small soliton is arranged to push a large one. The effective
potential acting on the soliton results of the deviation from
integrability. For the AL model (\ref{1}) that is integrable in
absence of randomness, the potential results of the randomness. For
the \emph{standard} NLS lattice (\ref{10}), the deviation from
integrability results of both randomness and discreteness, as it is
integrable in the continuum limit in absence of randomness. The
effect of the deviations results in the potential (\ref{11}) that
plays the role of $U(x)$ of (\ref{3.5}) found for the AL model
(\ref{1}). Therefore similar dynamics of solitons is expected in the
regime where the
effective potential description is valid.\\

Z.-Y. S. acknowledges the support in part at the Technion by a
fellowship of the Israel Council for Higher Education. This work was
partly supported by the Israel Science Foundation (ISF-1028), by the
US-Israel Binational Science Foundation (BSF-2010132), by the USA
National Science Foundation (NSF DMS 1201394) and by the Shlomo
Kaplansky academic chair.

\end{document}